\documentstyle[twoside,fleqn,npb,epsf]{article}
\newcommand{\Pom}{{\!I\!\!P}}
\newcommand{\GeV}{\mbox{ GeV}}
\def\Journal#1#2#3#4{{#1} {\bf #2}, #3 (#4)}
\def\NPB{{\em Nucl. Phys.} B}
\def\PLB{{\em Phys. Lett.}  B}

\def\PRD{{\em Phys. Rev.} D}

\def\ZPC{{\em Z. Phys.} C}

\title{Diffractive Interactions: Theory Summary\thanks{Talk given at
    the ${\rm 7^{th}}$ International Workshop on Deep Inelastic
    Scattering and QCD (DIS 99), Zeuthen, Germany, 19--23 April~1999,
    to appear in Nucl.\ Phys.\ B (Proc. Suppl.)}}

\author{M. Diehl\address{Deutsches Elektronen-Synchroton DESY, 22603
    Hamburg, Germany}}

\begin{document}
\begin{abstract}
  I review various theory issues in diffraction that have been
  presented and discussed in the working group, and a few points
  concerning the comparison of theory with data. Some common notation
  used in diffractive DIS is given in an appendix.
\end{abstract}

\maketitle

\section{Diffraction in DIS}

There has been an ongoing effort in the last years to describe
diffractive DIS within the framework of QCD. Progress has been made in
understanding the connections between different approaches, and we
have had presentations about modelling the nonperturbative input
needed in a QCD description.

Soper \cite{Soper} recalled that to leading twist accuracy, i.e.\ up
to corrections in powers of $1/Q^2$, the inclusive diffractive cross
section factorises into a hard photon-parton scattering subprocess and
diffractive parton distributions. This factorisation has been shown to
hold to all orders in perturbative QCD \cite{Collins} and may be
regarded on the same footing as the corresponding theorem for the
inclusive DIS cross section. Diffractive parton distributions are
defined through quark and gluon field operators in a similar way as
ordinary parton distributions or fragmentation functions; consequences
of this are that they are process independent, and that their
dependence on the factorisation scale is governed by the usual DGLAP
equations.

Two tasks follow from this: to test experimentally where and how well
these theory predictions are satisfied, and to measure the diffractive
parton densities as a source of information on the nonperturbative
physics at work in diffraction.

Two models for diffractive parton densities at a low starting scale
have been presented. Soper et al.\ \cite{Soper} have evaluated them
for a small-size hadron, coupling to a heavy quark-antiquark pair,
which can be calculated perturbatively. Under the hypothesis that the
basic features of the ans\-wer survive in the nonperturbative regime
relevant for a proton target the result is then compared with data on
$F_2^D$, and does indeed show the correct qualitative behaviour. In a
similar spirit Mueller's dipole approach to the BFKL pomeron, which is
based on heavy onium scattering, is being used by Peschanski et al.\ 
\cite{Peschanski} for the description of $F^{\phantom{.}}_2$ and
$F_2^D$.

Hebecker et al.\ \cite{Hebecker} have taken the opposite extreme of a
very large hadron and modelled diffractive and non-diffractive parton
distributions of the proton in their semiclassical approach. Note that
this description is originally formulated in a frame where the target
proton is at rest: the fast-moving $\gamma^*$ splits into $q\bar{q}$
or $q\bar{q}g$ partonic states which are scattered in the colour field
of the target. The expression of the amplitude obtained in this way
can however be re-interpreted in the Breit frame: to leading accuracy
in $1/Q^2$ it displays factorisation into a hard photon-parton
scattering and diffractive or non-diffractive parton densities,
including their QCD evolution. In a simple model for the colour field
of a large target Hebecker et al.\ obtain parton distributions in fair
agreement with the data on $F^{\phantom{.}}_2$ and $F_2^D$.

It is remarkable that two rather opposite model assumptions of a very
small and a very large hadron give results that have several
similarities. One is the large amount of gluons compared with quarks
in the diffractive distributions, and the other the behaviour of these
distributions at small and large parton momentum fraction $z$ with
respect to the Pomeron momentum $x_\Pom p$:
\begin{equation}
\frac{dq(z)}{dx_\Pom\, dt} \stackrel{z\to0}{\sim} \mbox{const} \, ,
\hspace{2em}
\frac{dq(z)}{dx_\Pom\, dt} \stackrel{z\to1}{\sim} (1-z)
\end{equation}
for quarks and
\begin{equation}
\frac{dg(z)}{dx_\Pom\, dt} \stackrel{z\to0}{\sim} \frac{1}{z} \, ,
\hspace{2em}
\frac{dg(z)}{dx_\Pom\, dt} \stackrel{z\to1}{\sim} (1-z)^2
\end{equation}
for gluons. Notice that such gluon distributions gently fall off as
$z\to 1$ and do not correspond to a ``super-hard'' gluon.

W\"usthoff \cite{Wusthoff} has pointed out that the wave function of a
$\gamma^*$ splitting into $q\bar{q}$ or $q\bar{q}g$ appears in these
calculations, as well as in the evaluation of two-gluon exchange by
Bartels et al.\ \cite{Bartels}, in earlier work by Nikolaev
\cite{Nikolaev-old}, and in the dipole approach to BFKL
\cite{Peschanski}. It is the idea underlying the BEKW parametrisation
\cite{Wusthoff,Royon-exp} that the kinematic factors provided by the
photon wave functions control the behaviour of $F_2^D$ at the
endpoints $\beta\to 0$ and $\beta\to 1$. For the leading twist part
this is related with the endpoint behaviour of the diffractive parton
densities; in particular a gluon density going like $(1-z)^{n}$ for
$z\to 1$ corresponds to a $(1-\beta)^{n+1}$ behaviour at $\beta\to 1$
for the contribution of boson-gluon fusion to $F_2^D$. It would be
interesting to understand in more detail to which extent the
perturbative physics of the photon wave function can account for the
$\beta$-dependence of $F_2^D$ and to which extent this dependence
reflects the nonperturbative dynamics of gluons in the proton.

In the comparison of theory with experiment the large-$z$ behaviour of
the diffractive gluon density still remains to be understood; a
reflection of this is the existence of two solutions in the BEKW fit
to the H1 data \cite{Royon-exp}, one corresponding to a very ``hard''
gluon, the other to a fairly ``soft'' one.

An important part of the programme to investigate diffractive parton
densities, and to constrain their shapes, is to look at other
processes where diffractive factorisation is expected to hold. With
densities obtained from an analysis of $F_2^D$ data it is indeed
possible to describe diffractive dijet production in $\gamma^* p$
collisions, and in $\gamma p$ collisions in the region where they are
dominated by the direct, pointlike component of the photon
\cite{Schilling}. 

What the implications are of the H1 and ZEUS data on diffractive charm
production \cite{charm,Whitmore} is too early to say and will have to
be clarified. It has been emphasised in the discussions that if this
process is compared with the results of two-gluon exchange
calculations then both the $c\bar{c}$ and $c\bar{c}g$ final states
must be taken into account, the latter being important at small
$\beta$. Note also that the $c\bar{c}$ final state is not included in
a description based on diffractive parton densities (unless one
introduces a diffractive charm quark distribution).

Further information can be expected from studies of the diffractive
final state. At this point it is important to remember that part of
the final state configurations is not included in the leading-twist
description with diffractive parton distributions. An example is a
$q\bar{q}$-pair with large relative transverse momentum, originating
from a longitudinally polarised $\gamma^*$. Its importance for the
analysis of $F_2^D$ at large $\beta$ has been stressed
\cite{Peschanski,Wusthoff}, in particular because of its influence on
the scaling violation pattern. Bartels \cite{Bartels} has reported on
work to calculate the $q\bar{q}g$ final state in the two-gluon
exchange picture for a wider part of phase space than where it is
known so far, namely for configurations where quark and antiquark do
not balance in transverse momentum and where the gluon is not
approximately collinear with the initial proton. One motivation of
this study is that for configurations with only high-$p_T$ partons in
the diffractive system one may expect a steeper energy dependence than
for the inclusive cross section as a manifestation of hard pomeron
dynamics.

Williams et al.\ \cite{Williams} have investigated the restrictions on
the diffractive system imposed by a rapidity gap cut in the HERA
frame, following earlier work by Ellis and Ross. They find that no
effect is to be expected with presently used values of $\eta_{\it
  max}$, but advocate to use data with stronger cuts as a means to
study the structure of the final state in a way that is sensitive to
the diffractive mechanism.

\section{Leading baryons in DIS}

The concept of diffractive factorisation can be extended to
non-diffractive production of a leading proton or neutron; in fact
diffractive parton densities are a special case of fracture functions
\cite{fracture}, which describe semi-inclusive particle production in
the target fragmentation region. The data on leading baryon production
\cite{leading} do indeed indicate that this factorisation is
satisfied, and in particular show that in DIS leading baryon
production is a leading twist phenomenon, as is diffraction. It is
further found that, when the cross section is integrated over a
certain range of momentum fraction for the leading baryon, its
dependence on Bjorken-$x$ follows that of the inclusive cross section,
which supports the idea of limiting fragmentation \cite{leading}.

\section{The energy dependence and the BFKL pomeron}

The same similarity in energy dependence has been observed some time
ago in the diffractive regime, and led Whitmore et al.\ to compare
diffractive quantities (integrated over a certain range in $x_\Pom$ at
fixed $x$) with inclusive ones at the level of parton densities
\cite{Whitmore}. It should also be noticed that a number of models can
actually make simultaneous predictions for $F_2^D$ and
$F^{\phantom{.}}_2$ \mbox{\cite{Peschanski,Hebecker,Wusthoff}}, and
one may hope that more will be learnt from confronting the dynamics of
inclusive and diffractive DIS.

While for the dependence of $F_2^D$ on $Q^2$, and to a lesser degree
on $\beta$, a number of predictions can be made in QCD and a certain
convergence between theory approaches has been achieved, it is fair to
say that the energy or $x_\Pom$-dependence is still poorly understood.

\begin{figure}
\begin{center}
  \leavevmode
  \epsfxsize=0.45\textwidth
 \epsfbox{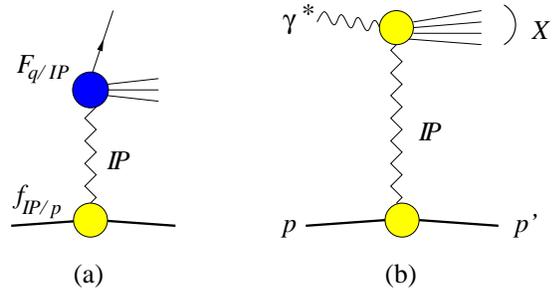}
 \vspace{0em}
\end{center}
\caption{\label{fig:regge} Two different types of Regge
  factorisation discussed in the text: (a) of a diffractive parton
  density and (b) of a diffractive scattering process.}
\end{figure}

Applying the ideas of Regge phenomenology to diffractive parton
densities one arrives at the Ingelman-Schlein proposal \cite{IS}. In
this scenario diffractive parton distributions factorise into a flux
factor $f_{\Pom/p}$ and parton distributions $F_{q,g/\Pom}$ of the
pomeron, cf.\ Fig.~\ref{fig:regge} (a),
\begin{equation} \label{Regge-fact}
  \frac{dq,dg(z,x_\Pom,t)}{dx_\Pom\, dt} = 
  f_{\Pom/p}(x_\Pom,t) \, F_{q,g/\Pom}(z,t) \, ,
\end{equation}
where the flux factor can be obtained from the Regge phenomenology of
soft hadronic reactions. A slightly more general ansatz with a sum
over contributions from the pomeron and various reggeons turns out to
work rather well as analyses of the data for $F_2^D$ and for leading
baryons show \cite{Royon-exp,leading}. As is well known the pomeron
intercept extracted in diffractive DIS is larger than the one found in
hadron-hadron scattering and in photoproduction. It remains to be
understood how such a Regge description can go together with the
observed similarity in the energy dependence of $F^{\phantom{.}}_2$
and $F_2^D$ mentioned above.

It may be worthwhile to notice that if at some factorisation scale
$Q_0^2$, large enough to use DGLAP evolution, the $x_\Pom$-dependence
of the diffractive parton distributions factorises as in
(\ref{Regge-fact}), then this dependence remains unchanged when one
evolves to higher $Q^2$, and with it the energy dependence of the
leading twist part of $F_2^D$ \cite{Soper,Buchmuller}. This is very
much in contrast to the $x$-dependence in $F_2$, which is made steeper
by evolution.

Yet another kind of factorisation is Regge factorisation (or $k_T$
factorisation in the context of perturbative QCD) of the entire
diffractive process $\gamma^* p \to X p'$ into pomeron exchange and
impact factors, describing the transitions $\gamma^*\to X$ and $p\to
p'$, cf.\ Fig.~\ref{fig:regge} (b). It should be remembered that this
factorisation goes beyond leading twist, as the $\gamma^*\to X$ impact
factor contains more than the leading power in $1/Q^2$. The role of a
large scale $Q^2$ here is to introduce perturbative QCD dynamics into
the process.

To probe the pomeron in a dynamical situation where it is as much
dominated by hard physics as possible, two types of "gold plated"
processes are being discussed: high energy $\gamma^* \gamma^*$
collisions, where the pomeron couples to a hard scale at both ends,
and diffraction at high $t$, where a large momentum is transferred
across the $t$-channel. 

Royon et al.\ \cite{Royon-the} have studied the total $\gamma^*
\gamma^*$ cross section in the dipole BFKL approach. Choosing similar
virtualities of the two photons should provide a means to separate
BFKL from DGLAP dynamics (which describes the evolution between two
different momentum scales), something which cannot be achieved in the
proton structure functions $F^{\phantom{.}}_2$ or $F_2^D$. It may be
worth noting that the leading order BFKL prediction for $\gamma^*
\gamma^*\to X$ is excluded by the L3 data \cite{Royon-the,Vogt}. In a
phenomenological estimate of NLO corrections Royon et al.\ find that
the effect of BFKL resummation over bare two-gluon exchange could be
seen at a linear collider, while the discriminating power of LEP2 is
marginal.

Theory aspects of high-$t$ diffraction in $ep$ or hadron-hadron
collisions have been discussed by Forshaw \cite{Forshaw}. He
emphasised that within the BFKL resummation even a moderately large
momentum transfer across the gluon ladder is very efficient in
suppressing the dangerous infrared regions in the phase space
integrations, where the results of perturbation theory become
doubtful. He also pointed out the virtues of large-$t$ diffractive
production of $J/\Psi$, light mesons or real photons, processes which
do not suffer from the difficulties of hadronisation effects and
rapidity gap survival encountered in forward high-$p_T$ jet
production, and according to estimations give observable event rates
in large parts of phase space.  Inclusive high-$t$ diffraction $\gamma
p \to XY$, defined by the largest rapidity separation in the event has
been discussed by Cox \cite{Cox}.

On the theoretical side progress has been reported in understanding
the structure of the BFKL pomeron. Kotsky \cite{Kotsky} has verified
that the impact factors occurring in the BFKL equation at NLO satisfy
a condition of self-consistency for the reggeisation of the gluon. At
the level of LO accuracy the subject of multi-pomeron couplings and
unitarity corrections has received much interest. Ewerz \cite{Ewerz}
recalled that in order to satisfy the unitarity bound the leading
$\log(1/x)$ approximation has to be relaxed to take into account
diagrams with more than 2 reggeised gluons in the $t$-channel.
Investigating the transitions between 2 and up to 6 gluons he found a
structure consistent with the unitarity of an underlying effective
field theory at high energy, based on conformal invariance, which
remains to be formulated. In the dipole approach Peschanski
\cite{Peschanski} has obtained several exact results for multi-pomeron
vertices, noting that the corresponding pomeron configurations may be
phenomenologically accessible in the triple Regge regime of $ep$
diffraction and for various types of gap-jet events at the Tevatron.
Gay Ducati \cite{Ducati} presented an evolution equation for $F_2$
which incorporates unitarity corrections and contains the
Gribov-Levin-Ryskin equation as a limiting case. This equation can be
derived within the dipole pomeron approach, where it corresponds to
parton recombination in the colour dipole cascade initiated by the
virtual photon.

\section{Diffraction in $p\bar{p}$ and $ep$ collisions}

The confrontation of rapidity gap events at the Tevatron with those at
HERA provides an opportunity to learn about the interplay between hard
and soft dynamics and about the transition from partons to hadrons in
reactions with a hard scale.

While in diffractive DIS there is a factorisation theorem for the
inclusive cross section and factorisation is expected to hold for hard
diffractive $\gamma p$ processes where they are dominated by the
pointlike component of the photon, there are theory arguments that in
hadron-hadron diffraction factorisation should break down due to
interactions between the spectator partons of the participating
hadrons.  This expectation is borne out in the comparison between HERA
and Tevatron data: the presentations by Whitmore \cite{Whitmore} and
the experiments \cite{Tevatron} consolidate previous statements that
with diffractive parton distributions which fit the inclusive $F_2^D$
and diffractive jet production data at HERA (and which are thus
required to contain a significant amount of gluons) predictions for
Tevatron processes come out far too big. There are also hints for
factorisation breaking in diffractive jet photoproduction in the
region of $x_\gamma$ where the hadronic component of the photon
becomes important \cite{Schilling}.

A way of quantifying the phenomenon of factorisation breaking is the
concept of gap survival probability, according to which a potential
rapidity gap left by a hard subprocess is filled by hadrons produced
in collisions between spectator partons. Gotsman \cite{Gotsman}
presented a model for the survival probability, where in particular a
strong dependence of gap survival on the total energy in the reaction
is found.

The experimental observation that the transverse energy spectra in
double diffractive, single diffractive and non-diffractive jet
production at the Tevatron look very similar \cite{Tevatron,Whitmore}
supports the picture that in all cases one has to do with the same
hard partonic subprocesses, and that it is mainly soft interactions
which may or may not destroy the rapidity gap, while not modifying the
large-$E_T$ spectrum of the jets.

A particular implementation of the idea that hard diffractive and
non-diffractive events can be described by a perturbative subprocess
and nontrivial dynamics of hadronisation which determines whether
there will be a rapidity gap or not, is the soft colour interactions
model. Ingelman \cite{Ingelman} showed that in this model a number of
processes both at the Tevatron and at HERA can be fairly well
described, without incurring the huge discrepancies in rates of the
factorisation ansatz. The physics assumption underlying soft colour
interactions is a rearrangement, before hadronisation, of the colour
strings between partons due to their interaction with a colour
background field. Ingelman further presented an alternative mechanism
based on re-interactions among the strings themselves, with the
hypothesis that these interactions tend to minimise the phase space
area ``swept out'' by the strings. This rather simple model is able to
give a reasonable description of $F_2^D$.

\section{Light meson production}

\begin{figure}
\begin{center}
  \leavevmode
  \epsfxsize=0.35\textwidth
 \epsfbox{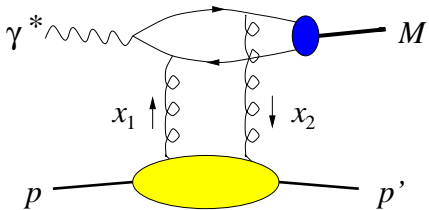}
 \vspace{0em}
\end{center}
\caption{\label{fig:meson} Leading-twist factorisation of meson
  production from a longitudinal photon into a skewed parton
  distribution in the proton, a hard scattering and the distribution
  amplitude of the meson.}
\end{figure}

Exclusive vector meson production has long been a major source of
information in diffractive physics. Within perturbative QCD it has
been shown \cite{factor} that in the Bjorken limit of large $Q^2$ at
fixed $x$ and $t$, and for longitudinal polarisation of the initial
photon, the amplitude for $\gamma^* p \to M p'$ factorises into a
skewed parton distribution in the proton, a hard parton scattering and
the distribution amplitude of the meson $M$, cf.\ 
Fig.~\ref{fig:meson}. The amplitude for transverse photons should be
power suppressed by $1/Q$. How far one is from the asymptotic regime
can thus in particular be studied with polarisation observables. The
data on the ratio $R = \sigma_L / \sigma_T$ of cross sections for
longitudinal and transverse photons have indicated for some time that
there is a substantial amount of transverse cross section even at a
$Q^2$ of $10 \GeV^2$ or more, and the measurements of the full decay
angular distributions \cite{rho} provide a wealth of information about
nonleading twist phenomena and the physics of the photon-$\rho$
transition in a perturbative regime. In this sense their importance
goes well beyond the statement that $s$-channel helicity is not
conserved to an accuracy better than some 10\%. The simplest
descriptions of the $\rho$, be it through a distribution amplitude,
where the relative transverse momentum between quark and antiquark is
integrated out, or as a nonrelativistic bound state where a
constituent $q\bar{q}$-pair equally shares the meson momentum, are
both too simple to describe this process beyond a 10\% accuracy, and
the data on the $\rho$ polarisation density matrix strongly indicate
that the inclusion of transverse momentum of the $q\bar{q}$-pair is
essential. Three implementations of this, using rather different
frameworks and physics assumptions, have been presented by Kirschner,
Nikolaev and Royen \cite{Kirschner,Nikolaev,Royen}. Within the present
experimental errors they can all account for the data, in particular
for the pattern of $s$-channel helicity violation, but they differ
among themselves to an extent that it may in the future be possible to
learn which are the adequate degrees of freedom in the transition from
$\gamma^*$ to $\rho$.

In diffractive $\rho$-production at large $t$, already mentioned in
connection with the search for the perturbative pomeron
\cite{Forshaw}, the photon-$\rho$ transition can again be studied in a
region where one should be able to describe it within perturbative
QCD. Also here polarisation observables can provide important insight
into the dynamics, and it will be interesting to understand the
results on the $\rho$ polarisation at high $t$ shown by Crittenden
\cite{Crittenden}.

Polyakov \cite{Polyakov} has shown that the leading twist description
of $\rho$-production can be extended to the production of a
$\pi^+\pi^-$-pair with invariant mass $M_{\pi\pi}^2\ll Q^2$, be it on
or off the $\rho$-peak. The corresponding generalised distribution
amplitude describes how a pion pair is formed out of a fast-moving
$q\bar{q}$-pair, and is related by crossing with the parton
distributions of the pion. It offers a new way to look at the
distribution amplitude of a resonance, and also shows that in the
factorising regime it is not necessary to extract a ``$\rho$-signal''
from the pion invariant mass spectrum in order to study, say, the
$x_\Pom$- and $t$-dependence of the process and the physics of skewed
parton distributions.

The cornerstone of understanding diffraction in QCD is the gluon
exchange picture, which has as a natural implication that an odderon
should exist as the negative charge conjugation partner of the
pomeron. No experimental evidence for this object exists so far.
Exclusive production of a pseudoscalar instead of a vector meson
offers a way to look for odderon exchange in $ep$ collisions. Using
the description of high-energy scattering developed by the Heidelberg
group, which accounts quite well for data on elastic hadron-hadron
scattering and exclusive $\rho$-production, Berger \cite{Berger} has
presented an estimate of the cross section for photoproduction of a
$\pi^0$, going along with proton dissociation. He found rates that
should make it possible to discover the odderon at HERA.

\section{Heavy meson production and skewed parton distributions}

Since the first data on exclusive $\Upsilon$ photoproduction have been
presented a year ago there have been improvements in the theory of
this process, for which at the time predictions were far below the
measured cross sections (while with the same model assumptions
$J/\Psi$-production could be described rather well). A number of
simplifying assumptions had to be refined, and the presentations by
McDermott and Teubner \cite{McDermott,Teubner} showed that while
differing in details both groups obtain cross sections in fair
agreement with the data. Points of debate are mainly the choice of
factorisation scale in the gluon distribution, and the question of how
good a nonrelativistic approximation is for the $\Upsilon$ wave
function (in addition, Teubner et al.\ also give a result based on
parton-hadron duality). An effect which increases the cross section
compared with the "naive" result is the inclusion of the real part of
the scattering amplitude, whose importance is related with the very
steep energy dependence of the skewed gluon distribution at the high
factorisation scale provided by the $\Upsilon$ mass.  Perhaps even
more spectacular is the effect of the ``skewedness'' in the gluon
distribution, i.e.\ the difference between the momentum fractions
$x_1$ and $x_2$ of the two exchanged gluons (cf.\ 
Fig.~\ref{fig:meson}), which is fixed at $x_1-x_2 = (M^2_V+Q^2) /W^2$
for the production of a meson with mass $M_V$. Estimations find that
due to the large mass of the $\Upsilon$ the effect of this asymmetry
amounts to a factor of 2 to 3 in the cross section, compared to just
approximating the skewed gluon distribution by the ordinary one, where
$x_1=x_2$.

Important theory progress has been made in the understanding of skewed
parton distributions, in particular concerning their evolution (cf.\ 
also the presentations in the spin working group of this workshop).
Martin \cite{Martin} has given arguments why for small $x_1$ and
$x_1-x_2$ the effect of the asymmetry between $x_1$ and $x_2$ at a low
factorisation scale becomes more and more washed out as one evolves to
large scales, so that the asymmetry there becomes increasingly
dominated by the dynamics of the evolution. One thus expects to obtain
a good approximation of skewed distributions at a high factorisation
scale by evolving them from a low scale, approximating the skewed
distributions at the starting scale by the ordinary ones in an
appropriate manner. Different ways of performing this approximation
have been presented by Martin and by Golec-Biernat \cite{Golec}. This
procedure then relates skewed and ordinary distributions in a
nontrivial but controlled way.

In the small-$x$ regime the measurement of skewed parton (mainly
gluon) distributions may in such a way be used to obtain information
on the ordinary gluon distribution. For larger $x$, where also the
asymmetry $x_1-x_2$ is larger, one can expect that the skewed
quantities (now mainly the quark distributions) will contain
nonperturbative information on the proton structure that cannot be
obtained from the ordinary ones. The kinematics at HERMES allows one
to study this regime, and the first comparison of exclusive
$\rho$-production data \cite{Borissov} with an estimate based on
skewed quark distributions is encouraging for the applicability of
this description at lower energies.

Freund \cite{Freund} has emphasised that the most direct information
on skewed distributions may be obtained in deeply virtual Compton
scattering, $\gamma^* p \to \gamma p$. On one hand there is no second
nonperturbative unknown like a meson wave function, and on the other
hand the interference of Compton scattering with the Bethe-Heitler
process in $e p \to e \gamma p$ offers a possibility to measure the
new distributions at \emph{amplitude} level. In particular the
different ways how the skewed distributions enter in the real and
imaginary parts of the Compton amplitude contains valuable
information.  The interference term is accessible through an azimuthal
asymmetry, and even more directly through the asymmetries in the beam
lepton charge ($e^+$ vs.\ $e^-$) or polarisation.

With $\Upsilon$-production providing probably the first evidence for
nontrivial effects of skewedness, it can be hoped that future data on
various processes will enable us to make use of skewed parton
distributions as an additional tool to study hadron structure.

\section*{Acknowledgements}
It is a pleasure to thank many participants of our working group for
discussions, my co-conveners for the pleasant collaboration, and the
organisers for the smooth running of this workshop. I am grateful to
M.  McDermott for making available the ``diffractive DIS convention
summary'' \cite{DIS98}, on which the present appendix is
based. Special thanks are due to W. Buchm\"uller for many discussions,
and to T. Teubner for a careful reading of the manuscript.

\appendix
\section*{Appendix}

\begin{figure}
\begin{center}
  \leavevmode
  \epsfxsize=0.45\textwidth
 \epsfbox{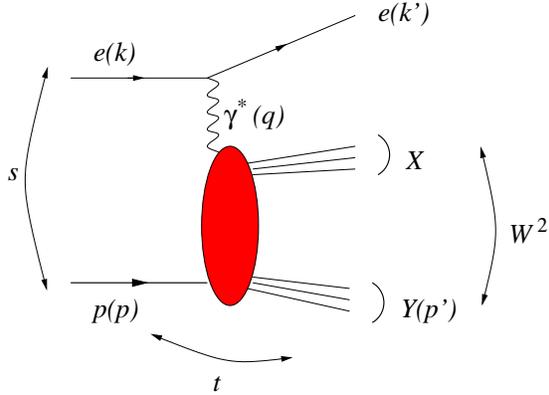}
 \vspace{0em}
\end{center}
\caption{\label{fig:kin} Definition of some kinematic variables in a
  generic diffractive process $ep \to eXY$. It includes the special
  cases when $Y$ is a proton or $X$ a vector meson. Between the
  hadronic systems $X$ and $Y$ there is a gap in rapidity.}
\end{figure}

This appendix gives some commonly used notation in diffractive DIS.
Four-momenta are defined in Fig.~\ref{fig:kin}.

\begin{list}{$\bullet$}{\addtolength{\leftmargin}{-1em}}
\item General DIS variables:
\begin{eqnarray*}
Q^2 & = & - q^2      =  - (k - k')^2 \nonumber \\
W^2 & = & (p + q)^2                  \nonumber \\
x   & = & \frac{Q^2}{2 p\cdot q } = \frac{Q^2}{W^2 + Q^2 - m_p^2} 
                                     \nonumber \\
s   & = & (p + k)^2                  \nonumber \\
y   & = & \frac{q\cdot p}{k\cdot p} = \frac{W^2 + Q^2 - m_p^2}{s-m_p^2} 
\end{eqnarray*}
\item Diffractive DIS variables:
\begin{eqnarray*}
t & = &  (p - p')^2          \nonumber \\
M_X^2  & = & (p - p' + q)^2  \nonumber \\
M_Y^2  & = &  p'^2           \nonumber \\
x_\Pom & = & \frac{(p-p')\cdot q}{p\cdot q} 
            = \frac{M_X^2 + Q^2 - t}{W^2 + Q^2 - m_p^2} 
                                  \nonumber \\
\beta  & = & \frac{Q^2}{2(p-p')\cdot q}  =  \frac{Q^2}{M_X^2 + Q^2 -t}
            = \frac{x}{x_\Pom}
\end{eqnarray*}
It is also common to write $\xi$ instead of $x_\Pom$.
\item Diffractive structure functions:
\begin{eqnarray*}
\lefteqn{\frac{d^{4}\sigma(ep \rightarrow eXY)}{dx\, dQ^{2}\,
    dx_\Pom\, dt} = \frac{4\pi\alpha_{em}^2}{xQ^4}} \nonumber \\
&& 
\times \left[ \Big(1-y+ {y^2\over2}\Big)\, F_2^{D(4)}
      - {y^2\over2}\, F_L^{D(4)}  \right]
\end{eqnarray*}
\begin{displaymath}
F_2^{D(4)} = F_T^{D(4)} + F_L^{D(4)}
\end{displaymath}
\item $t$-integrated diffractive structure functions:
\begin{eqnarray*}
\lefteqn{F_{i}^{D(3)}(x_{\Pom},\beta,Q^{2}) =} \\
&& \int_{|t|_{\mathit{min}}}^{|t|_{\mathit{max}}} 
d|t|\, F_{i}^{D(4)}(x_{\Pom},\beta,Q^{2},t)
\end{eqnarray*}
with $i = 2,T,L$. Here $|t|_{\mathit{min}}$ is the lower kinematic
limit of $|t|$ and $|t|_{\mathit{max}}$ has to be specified.
\end{list}

\end{document}